\documentclass[fleqn,11pt]{wlscirep}

\usepackage{epstopdf}

\title{Brain structural connectivity atrophy in Alzheimer's disease}

\author[1,2]{Nicola Amoroso}
\author[1,2,*]{Marianna La Rocca}
\author[3]{Stefania  Bruno}
\author[1,2]{Tommaso Maggipinto}
\author[2]{Alfonso Monaco}
\author[1,2]{Roberto Bellotti**}
\author[2]{Sabina Tangaro**}
\author[ ]{for the Alzheimer's Disease Neuroimaging Initiative\footnote[4]{Data used in preparation of this article were obtained from the Alzheimer's Disease Neuroimaging Initiative (ADNI) database (adni.loni.usc.edu). As such, the investigators within the ADNI contributed to the design and implementation of ADNI and/or provided data but did not participate in analysis or writing of this report. A complete listing of ADNI investigators can be found at: \url{http://adni.loni.usc.edu/wp-content/uploads/how_to_apply/ADNI_Acknowledgement_List.pdf}.}
}

\affil[1]{Universit\`a degli studi di Bari, Dipartimento Interateneo di Fisica, Italy.}
\affil[2]{Istituto Nazionale di Fisica Nucleare, sez. di Bari, Italy.}
\affil[3]{Blackheath Brain Injury Rehabilitation Centre, London, UK.}
\affil[*]{email to: marianna.larocca@ba.infn.it}
\affil[**]{RB and ST: equal last-author contribution}

\begin{abstract}
Analysis and quantification of brain structural changes, using Magnetic resonance imaging (MRI), are increasingly used to define novel biomarkers of brain pathologies, such as Alzheimer's disease (AD). Network-based models of the brain have shown that both local and global topological properties can reveal patterns of disease propagation. On the other hand, intra-subject descriptions cannot exploit the whole information context, accessible through inter-subject comparisons. To address this, we developed a novel approach, which models brain structural connectivity atrophy with a multiplex network and summarizes it within a classification score. On an independent dataset multiplex networks were able to correctly segregate, from normal controls (NC), AD patients and subjects with mild cognitive impairment that will convert to AD (cMCI) with an accuracy of, respectively, $0.86 \pm 0.01$ and $0.84 \pm 0.01$. The model also shows that illness effects are maximally detected by parceling the brain in equal volumes of $3000$ mm\textsuperscript{3} ("patches"), without any \textit{a priori} segmentation based on anatomical features.  A direct comparison to standard voxel-based morphometry on the same dataset showed that the multiplex network approach had higher sensitivity. This method is general and can have twofold potential applications: providing a reliable tool for clinical trials and a disease signature of neurodegenerative pathologies.
\end{abstract}
\begin{document}

\flushbottom
\maketitle
%
%
\thispagestyle{empty}


\section*{Introduction}

Alzheimer's disease (AD) is a progressive, neurodegenerative disease accounting for most cases of dementia after the age of  $65$. It is expected that over $115$ million people will develop AD by $2050$ \cite{prince2014world}. Illness related brain changes can be detected \textit {in vivo} with magnetic resonance imaging (MRI) and neuroimaging has been playing an increasingly important role for the diagnosis of neurodegenerative disorders \cite{dubois2014advancing,bron2015standardized} to the extent that it has been incorporated in the diagnostic criteria for AD \cite{McKhann2011Thediagnosis}. It is now accepted that the neurodegenerative cascade in AD begins in the brain years, decades even, before the clinical and radiological manifestations of the illness. The dementia is preceded by a prodromal phase of mild cognitive impairment \cite{albert2011diagnosis}, and this, in turn, by a pre-clinical phase \cite{sperling2011toward} of variable duration. Understanding the biological changes, occurring in these early phases, is of paramount importance, as it would open a window of opportunity for future disease-modifying treatments. While it is clear that neurodegeneration in AD occurs in a rather stereotyped fashion in the majority of cases \cite{west1994differences,perl2010neuropathology}, it is not known exactly what drives the propagation of the disease within an individual, and what is behind the variations in the patterns of atrophy between individuals. To which extent neurodegeneration propagates through anatomical contiguity, or through preferential pathways of structural or functional connectivity is yet to be clarified, but network-based approaches can yield a better understanding of these phenomena.

MRI can provide significant information on large-scale topological organization of the brain \cite{yao2010abnormal,alexander2012anatomical,tijms2013alzheimer} and graph theory has been widely used to study both functional and structural connectivity \cite{cciftcci2011minimum,stam2009graph,de2012disrupted,tijms2013alzheimer} for AD and, more in general, for neurodegenerative disease characterization \cite{greicius2012neuroimaging}. These studies reported altered local and global graph properties in AD, supporting the clinical relevance of brain networks, especially within group-wise association studies \cite{crossley2014hubs,daianu2015rich}. Up to now the different graph theory strategies used to model and describe the brain \cite{bullmore2011brain, bullmore2009complex,tijms2013alzheimer,he2008structural,cciftcci2011minimum,stam2009graph,sporns2013network} have been based on two distinct approaches \cite{suk2014hierarchical}: (i) voxel-wise  and (ii) region of interest analyses. We propose here a novel approach combining the interpretability of a voxel-wise description, without its intrinsic computational burden and noise sensitivity \cite{davatzikos2004voxel}, and the robustness of region of interest methods, avoiding \textit{a priori} assumptions in terms of disease effects or segmentation accuracy \cite{amoroso2015hippocampal}. In addition, as brain disease has often a diffuse effect, affecting multiple voxels, but not necessarily corresponding to entire anatomical structures, the proposed approach has the potential to better suit the description of pathological changes in the brain, reflecting biological variability.

Specifically for network science, recent studies have investigated the limitations of traditional approaches to describe real systems \cite{boccaletti2014structure,lee2012correlated,mucha2010community} and have pointed out that context information plays a fundamental role. Here we introduce an approach that segments the whole brain in rectangular boxes, from now onward referred to as ``patches'', representing the nodes of a network modeling each subject. The network connections are pairwise Pearson's correlations between the nodes. This is aimed at investigating graph-based properties from the different perspective of multiplex networks (from now onward also multiplexes) and introducing a novel concept, namely \textit{structural connectivity atrophy}. Multiplexes are multi-layer systems with a fixed number of nodes that can be linked in different interacting layers, to investigate inter-subject characterization, rather than group-wise differences. In this study multiplex-based features are exploited to efficiently model AD-related atrophy patterns; these faetures are then used within a random forest classifier to correctly segregate normal controls (NC) from AD patients and subjects with mild cognitive impairment that will convert to AD (cMCI). We demonstrate how a structural connectivity atrophy can be used to describe inter-subject variability relating it to the emergence of statistically significant AD patterns altering the topological organization of the brain.

\section*{Results}


\subsection*{Scale selection and informative content}
\label{sec:scale}

\begin{figure}[!htbp]
\centering
\includegraphics[scale=0.35]{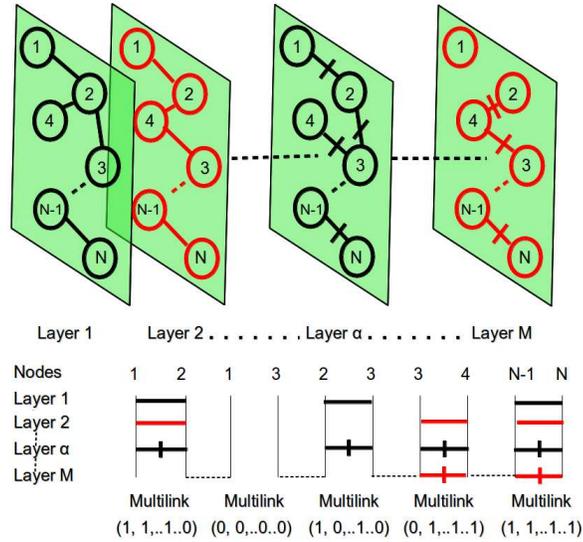}
\caption{At the top: the multiplex network with $M$ layers, representing a study subject, and $N$ nodes, representing brain patches. At the bottom: the representation of multi-links for the different pairs of network nodes. Within each layer different nodes can be connected with a link and a specific weight. This context information is then used to detect different patterns.}
\label{fig:multiplex}
\end{figure}

Graph theory provides tools to concisely quantify the properties of complex networks that describe interrelationships (represented by edges) between the objects of interest (represented by nodes). In this work, for each subject, and thus for each MRI brain scan, we built a weighted undirected network whose $N$ nodes were rectangular brain patches and whose connections were defined by measuring their pairwise Pearson's correlation, see Figure \ref{fig:multiplex} for a pictorial representation.


In order to detect structural local changes of the brain, we considered the strength $s$ of each patch, measuring the intensity of its connections. Strength does not take into account if the number of connections is preserved, thus we included in our representation the nodal degree $k$ which is the number of existing connections and the inverse participation ratio Y evaluating how unevenly the weights of the links of the node are distributed \cite{menichetti2014weighted}. To capture inter-subject variations, we introduced the conditional values of strength $s(k)$ and inverse participation $Y(k)$. Finally, to capture intra-subjects changes, we considered the distribution degrees $k'$ of the whole cohort and analogously defined the quantities $s(k')$ and $Y(k')$. In conclusion we obtained a 8-dimensional feature representation for each node.

There is no \textit{a priori} knowledge determining the dimension of patches and, therefore, the number of nodes $N$ that should be used to model a brain with a network. Thus, we firstly investigated if a privileged dimension existed in order to maximize classification accuracy to distinguish controls and patients. To this aim, we used a first random forest classifier for feature selection and a second random forest to summarize network measures in a unique network atrophy score, within a  $5$-fold cross-validation framework. This score allowed subjects' classification. Experimental results on a mixed cohort of $38$ AD and $29$ NC, namely $\mathcal{D}_{train}$, showed an accurate and stable classification performance within the $[2250,3200]$ voxel (mm\textsuperscript{3}) range, corresponding approximately to $500$ patches, with variations lower than $5\%$. 
This range established the optimal patch dimension for the NC-AD classification, best results were obtained with $N = 549$ patches. In addition, we performed a second classification test using structural morphological features obtained by FreeSurfer \cite{fischl2012freesurfer}, instead of multiplex features. The accuracy of the proposed methodology was on average $0.88$ with a $0.01$ standard error (sensitivity and a specificity respectively $0.90 \pm 0.01$ and $0.88 \pm 0.02$) while with FreeSurfer features we obtained $0.83 \pm 0.01$ (sensitvity $0.86 \pm 0.01$ and specificity $0.79 \pm 0.01$). This result confirms the effectiveness of the multiplex characterization and the possibility to use our framework to discriminate controls from patients using only MRI data.

\subsection*{Anatomical characterization}
\label{sec:clinic}

Once the optimal dimension of multiplex network had been fixed we investigated on $\mathcal{D}_{train}$ the most significant features according to their classification importance and consequently the related anatomical districts. The reason was twofold: reduce the data dimensionality and gain clinical insight.

%

\begin{figure}[!htbp]
\centering
\includegraphics[scale=.6]{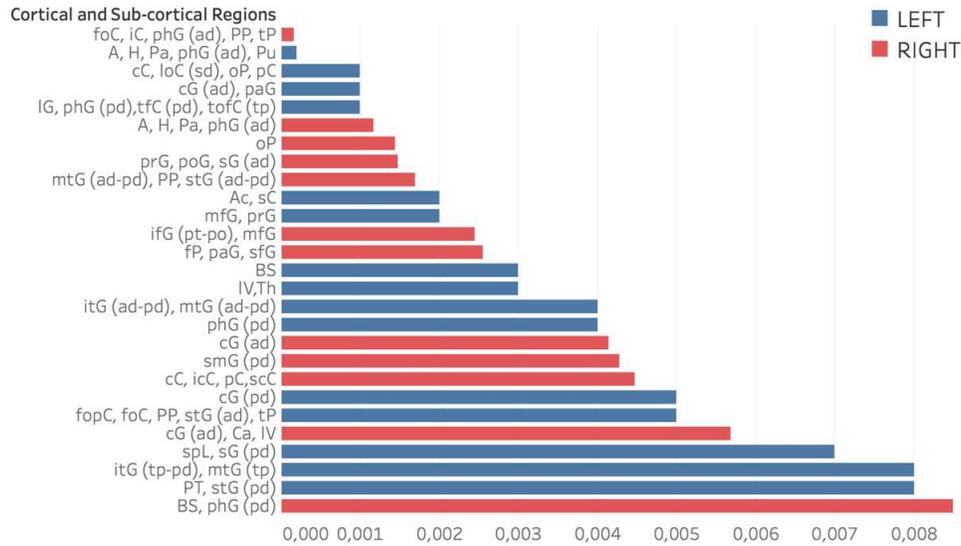} 
\caption{Regions related to AD in order of significance. \tiny  Accumbens (Ac), Amygdala (A), Brain-Stem (BS), Caudate (Ca), Cingulate Gyrus (cG) anterior division (ad), Cuneal Cortex (cC), Frontal Operculum and Orbital Cortex (fopC) and (foC), Frontal Pole (fP), Hippocampus (H), Inferior Frontal Gyrus (ifG) pars opercularis and pars triangularis (po) and (pt), Inferior Temporal Gyrus (itG) anterior division and temporoccipital part (tp), Insular Cortex (iC), Intracalcarine Cortex (icC), Lateral Occipital Cortex (loC) superior division (sd), Lateral Ventrical (lV), Lingual Gyrus (lG), Middle Frontal and Temporal Gyrus (mfG) and (mtG), Occipital Pole (oP), Pallidum (Pa), Paracingulate and Parahippocampal Gyrus (paG) and (phG), Planum Polare and Temporale (PP) and (PT). Postcentral and Precentral Gyrus (poG) and (prG), Precuneous Coretx (pC), Putamen (Pu), Subcallosal Cortex (sC), Superior Frontal Gyrus (sfG), Superior Parietal Lobule (spL), Superior Temporal Gyrus (stG), Supracalcarine Cortex (scC), Supramarginal Gyrus (sG), Temporal Fusiform and Temporal Occipital Fusiform Cortex (tfC) and (tofC), Temporal Pole (tP), Thalamus (Th). In parentheses: anterior, posterior and superior division (ad,pd,sd) and temporooccipital part (tp).}
\label{fig:ROI}
\end{figure}

Starting from the initial network of $549$ nodes, for each subject we determined which regions were able to effectively distinguish controls from patients, thus outlining $32$ significant patches, $18$ ($\sim 56\%$) in the left hemisphere and $14$ in the right; these regions included $27$ different cortical and sub-cortical regions listed in Figure \ref{fig:ROI} in order of significance. Further details about significance measures for both features and anatomical districts are provided in Methods section. As a region can be included in different patches (provided at least one of its voxels belongs to the considered patch), only most significant p-value entries are reported.

\begin{figure}[!htbp]
\centering
\includegraphics[scale=0.45]{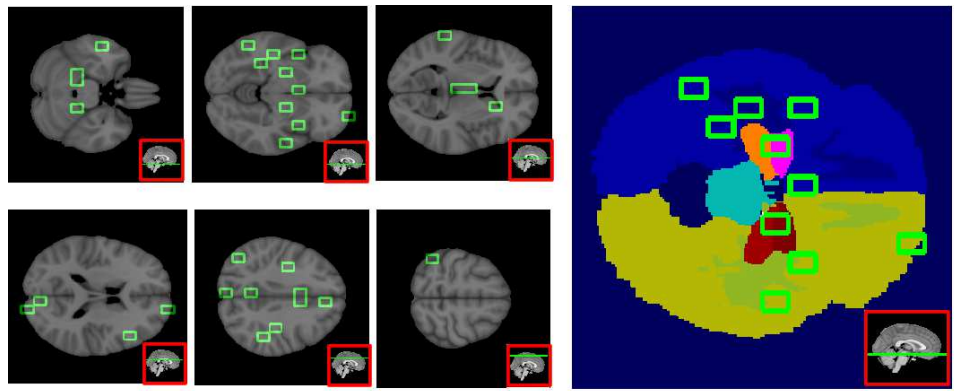} 
\caption{This figure shows six axial planes (left panel) with the significant patches outlined in green (p-value $< 0.01$), and on the right, the Harvard-Oxford Atlas used for the patch anatomical localization.}
\label{fig:anatPatch}
\end{figure}

In Figure \ref{fig:anatPatch} some representative brain axial planes are shown, as well as the Harvard-Oxford atlas \cite{desikan2006an} we used for this assessment.

\noindent 
In the left hemisphere, patches corresponding to amygdala, hippocampus, para-hippocampal gyrus, pallidum and putamen showed the strongest association to AD (p $= 0.0001$). For cingulate and para-cingulate giri, pre-cuneus, cuneus, and occipital cortex p $= 0.001$. Other significant patches (p $= 0.002$) were located in middle frontal gyrus and pre-frontal gyrus, nucleus accumbens, brain stem and thalamus.

On the right, p $= 0.0001$ for orbito-frontal cortex, insular cortex, prarahippocampal gyrus, planum polare and planum temporale; p $= 0.001$ for the parahippocampal-amygdalar complex, occipital pole, pre- and post-central gyri, supramarginal gyrus, middle and superior temporal gyri; p = $0.002$ for inferior, middle and superior frontal gyri, frontal pole, and paracingulate gyrus. Interestingly, the right frontal lobe involvement was more evident.

\subsection*{Multiplex Networks vs Voxel Based Morphometry}
\label{sec:lat_asy}

In order to establish if this new approach may offer any advantages over existing widely used methods, we analyzed the same data set with Voxel Based Morphometry (VBM) \cite{ashburner2000voxel}.

\begin{figure}[!htbp]
\centering
\includegraphics[scale=.45]{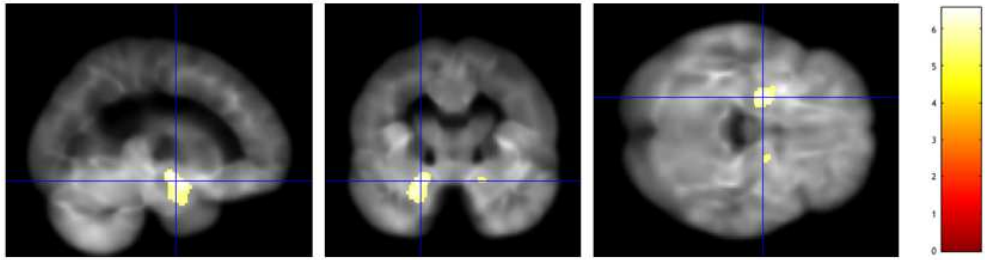} 
\caption{A voxel based morphometry analysis shows bilateral areas of significantly reduced grey matter density in patients with AD, in medial temporal lobe structures, such as hippocampus and amygdala, more prominent on the left as expected.}
\label{fig:spm}
\end{figure}

We followed the standard prescription for VBM with the publicly available SPM 12 suite\footnote{http://www.fil.ion.ucl.ac.uk/spm/software/spm12/}. Firstly, a segmentation of brain tissues was performed, followed by non-linear normalization with the SPM tool DARTEL to create a study specific template. Secondly, we performed a smoothing with an isotropic Gaussian filter with a full width at half maximum of $8$ mm. Lastly, a two-sample analysis was performed with a $t$ statistics to investigate significant group-wise differences in atrophy between NC and AD on training subjects. Significant voxels, with $5\%$ family-wise correction, are represented in Figure \ref{fig:spm}.

The VBM analysis showed significant reduction in grey matter density only in bilateral peri-hippocampal regions,  more prominent of the left. Compared to the proposed methodology, able to detect $32$ significant regions, VBM showed a largely decreased sensitivity. Since the VBM analysis confirmed that left-sided changes were more prominent, two dedicated tests were carried out to further explore the lateralization. Firstly, we used only multiplex features inherent to the left (right) hemisphere and trained the classification models. We found that left patches were able to discriminate NC from AD patients with an accuracy of $0.87 \pm 0.01$ while right hemisphere features were able to reach the accuracy value $0.85 \pm 0.01$. Left hemisphere remained responsible for a greater part of the overall information of the multiplex framework, which was $0.88 \pm 0.01$. 

However, each patch summarizes a network of interrelationships with other patches independently from its spatial collocation. As an example, the strength of a node denotes the sum of its connections, the fact that a node of the left hemisphere is significantly related to AD does not prevent its strength to be the result of its correlation with the right hemisphere. As a consequence, a second test was performed. We built the multiplexes of left and right hemispheres separately, thus disregarding one hemisphere. Classification accuracy for NC-AD when using left (right) multiplex was $0.83 \pm 0.01$ ($0.81 \pm 0.01$), thus confirming a greater involvement of the left hemisphere but also signaling a definite deterioration of the information content if compared with the whole brain multiplex. 

\subsection*{Robustness and generalization}
\label{sec:robust}

\begin{figure}[!htbp]
\centering
\includegraphics[angle=270,scale=.35]{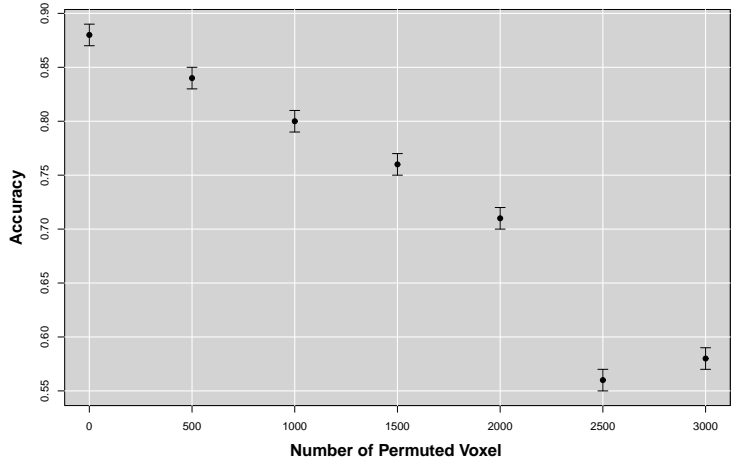} 
\caption{Accuracy varying with the number of permuted voxel within a patch. Classification performance decreased as the number of shuffled voxels was increased. Noticeably, a drastic drop was observed when the shuffle reached values of about $2500\sim3000$ voxels.}
\label{fig:perm_voxel}
\end{figure}

To investigate if classification performance based on netowrk atrophy was related to the random permutation of voxels inside a patch, we firstly shuffled a varying number of voxel within each patch, while keeping the patch decomposition stable, thus affecting the Pearson's correlation pairwise measurement. Then we measured the classification accuracy. The training results are presented in Figure \ref{fig:perm_voxel}.

The test was repeated $100$ times increasing the size of the shuffle by $500$ voxels at the time. It could be noticed that for small variations, under $1000$ voxels, performance did not suffer a significant deterioration; but with $2500$ voxel permutation a drastic drop of the performance was observed, a value comparable with the dimensional scale determined in section \ref{sec:scale}.

To further assess the method robustness we also performed a classical non-parametric statistical permutation test. This consisted in the permutation of the clinical labels of each subject belonging to $\mathcal{D}_{train}$. We performed $1000$ random permutation and observed  that the classification accuracy was $0.50 \pm 0.05$.



Training set randomization effectively established that the multiplex framework was able to model a significant structure in the $\mathcal{D}_{train}$ data between the multiplex features and the clinical label. Moreover, given the normality of the performance distribution obtained by permuting the labels, it was possible to assign a p-value to the performance obtained without permutations. The result showed that the multiplex model was able to identify a significant ($p<.001$) class structure within the $\mathcal{D}_{train}$ data. Otherwise, it would not have been possible to reject the null hypothesis underlying this test, \textit{i. e.} that labels and features were independent, so that in fact no difference really existed between the classes. 

As a further assessment, we performed a binary classification on an independent test set $\mathcal{D}_{test}$, incuding $48$ AD, $52$ NC and $48$ cMCI, for the NC-AD and NC-cMCI cases. The analysis was repeated using $100$ bootstrapped $\mathcal{D}_{test}$ sets to provide a measurement of the performance uncertainty. We found in terms of accuracy respectively $0.86 \pm 0.01$  and  $0.84 \pm 0.01$. The respective specificity were $0.74 \pm 0.01$ and $0.72 \pm 0.01$, while sensitivity reached higher values for both cases: $0.96 \pm 0.01$ and $0.94 \pm 0.01$. Remarkably, the NC-cMCI classification performance compared well with NC-AD classification confirming the method reliability and its informative content. 

The small, but significant, performance deterioration (training accuracy was $0.88 \pm 0.01$, see section \ref{sec:scale}) could be expected, mainly because even if the test perturbation of the training multiplex was considered small, it should not be completely neglected. The implementation of larger training sets could in principle mitigate this effect. 

%
%
%
%




 

\section*{Discussion}
\label{sec:disc}

The proposed approach aims at modeling  brain atrophy in AD through  inter-subject multiplex networks whose nodes are represented by brain patches and edges by pairwise Pearson's correlations. Basing on multiplex features we introduced a network atrophy score. This score allowed a robust classification performance over a broad range of Pearson's correlations and with the use of other similarity metrics (see sections ``Threshold Assessment'' and ``Similarity Metric Study" of the Supplementary Materials).


An optimal volume size for the detection of AD effects, maximizing the informative content of the multiplex, was identified as ranging from $2250$ to $3200$ mm\textsuperscript{3}. it is worthwhile to note that this range was comparable with the size of several brain structures related to AD, such as the hippocampus, therefore suggesting it as an ideal multiplex dimension for AD characterization.



The high sensitivity of the method in the detection of illness related brain changes was demonstrated by the number of regions, \textcolor{blue}{$32$}, that were identified as significantly associated with AD. The detected regions comprised hippocampus and para-hippocampal-amygdalar complex, pallidum and putamen, cingulate and paracingulate giri, pre-cuneus, cuneus, and occipital cortex, middle frontal gyrus, pre-central gyrus, accumbens, sub-callosal cortex and brain stem.

The cingulate cortex early involvement in AD pathology, has been amply demonstrated by functional and structural studies \cite{minoshima1997metabolic,bailly2015precuneus}. The same is true for posterior areas, such as cuneus and pre-cuneus, also known to be affected by the illness in early stages \cite{baron2001vivo,bailly2015precuneus}. As to the involvement of subcortical gray matter in AD, this has also been recognized, and shown to correlate with cognitive impairment \cite{de2008strongly}. Volume loss of the nucleus accumbens was found to increase the risk of progression from MCI to AD \cite{yi2015relation}. The brain stem is a key area in the early pathophysiology of Parkinson's disease, another common neurodegenerative disorder, and alterations of the brain stem in AD have been shown both in vivo \cite{lee2015brainstem} and post-mortem \cite{simic2009does}. 

It was striking how VBM on the same data set was able to detect only atrophy of the perihippocampal regions. The method here described seems more sensitive than standard VBM \cite{good2002voxel}, while studies adopting advanced VBM methodologies have also shown better results \cite{karas2003comprehensive}. The method outlined the involvement of $32$ significant brain regions, but only $22$ concerned single-layer measures;  thus, the multiplex model thus allowed a consistent increment ($+46\%$) in sensitivity. The results also confirmed asymmetry in the spatial distribution of significant patches, mostly located in the left hemisphere, in keeping with several other studies \cite{fennema2009structural,derflinger2011grey}. This asymmetry has a direct effect on the informative content (see ``Left/Right Characterization'' of the Supplementary Materials).


As to the application of this methodology to disease classification studies, we evaluated the method robustness on an independent set $\mathcal{D}_{test}$ and confirmed its reliability for discriminating both AD patients and cMCI subjects from controls. Classification performances are accurate, it should be noticed the obtained results are comparable with recent classification-focused studies \cite{moradi2015machine,salvatore2015magnetic,bron2015standardized}. To provide a diagnosis support system, although results are encouraging in this sense, was not the end goal of this work; however, the multiplex model is able to efficiently capture diseased patterns and inter-subject variability thanks to the specific multi-layer features this model can introduces. An even more refined classification could have been achieved including, as suggested by our previous works, structural features \cite{amoroso2014miccai} or longitudinal information \cite{chincarini2016integrating}. In addition, the method has great versatility and lends itself to a variety of purposes, including the identification of ``disease signature'' for more anatomically heterogeneous forms of neurodegenerative disorder, such as tauophathies or synucleinopathies, where the model could be enriched with additional clinical or genetic data.


\section*{Conclusion}
\label{sec:conc}

In this paper we propose a novel approach based on multiplex networks to characterize brain structural variations related to AD. We investigated the information content provided by multiplex networks and showed that they produce an accurate modeling of the disease.

We demonstrated how this framework is able to provide a robust method for AD characterization: (i) it shows the existence of an optimal scale for the description of disease effects of $[2250,3200]$ voxels, which reflects the size of brain structures relevant in AD, such as the hippocampi. (ii) The method does not require any \textit{a priori} human expert segmentation and correctly identifies cerebral region significantly related to AD. It also confirms that AD pathology is more prominent in the left hemisphere. (iii) Multiplex networks are a robust and effective method to describe disease patterns. 
Multiplex-based features allow, on the independent test set $\mathcal{D}_{test}$, the accurate classification of AD patients, with an accuracy of $0.86 \pm 0.01$, and cMCI subjects, with an accuracy of $0.84 \pm 0.01$, from NC subjects. 

The information content provided by multiplex characterization was able to efficiently detect disease patterns.  Also, the method is very suitable to application to longitudinal studies, ideally in association with functional imaging, to improve our understanding of the different patterns of neurodegeneration in different diseases. The impact of variables such as the degree of atrophy, disease duration, site or scanner type could also be investigated in further studies.


\section*{Methods}
\label{sec:mats}

\subsection*{Subjects}
\label{sec:subjs}
In this work we used a training a set $\mathcal{D}_{train}$ of $67$ T1 MRI scans, composed of 29 normal controls (NC) and 38 AD subjects, from the Alzheimer's Disease Neuroimaging Initiative (ADNI). These subjects belonged to a larger benchmark dataset selected in order to obtain a compact yet representative sample of ADNI \cite{boccardi2015training}; the dataset included also MCI subjects, which were excluded as not relevant to this study. We also employed an independent test set of $148$ subjects $\mathcal{D}_{test}$, composed by 52 NC, 48 AD and 48 subjects with mild cognitive impairment converting to AD (cMCI). $\mathcal{D}_{test}$ subjects were randomly chosen within the whole ADNI in order to match the demographic characteristics of training subjects. The training sample ($67$) and the test sample ($148$) are of sufficient size for the construction of robust classification models \cite{mukherjee2003estimating,beleites2013sample}. cMCI subjects had converted to AD in a time range of $[30,108]$ months subsequent to the initial assessment. All $215$ participants underwent whole-brain MRI at 34 different sites. Both $1.5$ T and $3.0$ T scans were included in $\mathcal{D}_{train}$ and $\mathcal{D}_{test}$.

ADNI images consisted of MPRAGE MRI brain scans, which were normalized with the MNI152 brain template of size of $197\ \times\ 233\ \times\ 189$ mm\textsuperscript{3} and resolution of $1 \times 1 \times 1$ mm\textsuperscript{3}; as a consequence voxels and mm\textsuperscript{3} can be interchangeably used. Clinical and demographic information, including the Mini Mental State Examination (MMSE) score, age, years of education and sex for the $\mathcal{D}_{train}$ and $\mathcal{D}_{test}$ is detailed in Supplementary Materials \ref{sec:dem}.

\subsection*{Multiplex network construction}
\label{sec:multi}

\begin{figure}[!htbp]
\centering
\includegraphics[scale=.45]{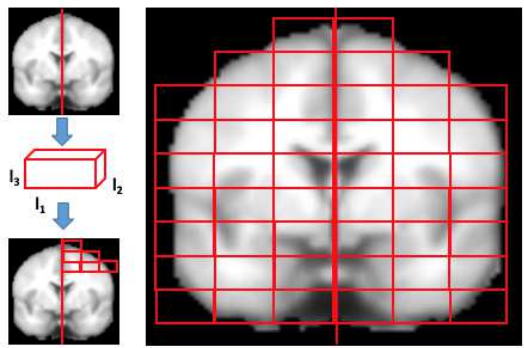}
\caption{The figure qualitatively shows how MRI brain scans are segmented in rectangular patches of dimensions $l_1 \times l_2 \times l_3$. Firstly, the brains normalized to MNI152 template are divided in left and right hemispheres using the medial longitudinal fissure, then the patch dimensions are set and finally the brain is segmented. Only patches overlapping the brain for at least the $10\%$ of their content are kept, others are discarded.}
\label{fig:figure_patches}
\end{figure}

A standard image processing procedure was carried out with the Oxford FMRIB library FSL \cite{jenkinson2012fsl}. Firstly, MRI scan intensity differences were normalized, then intra-cranial regions were extracted with the FSL Brain Extraction Tool (BET). Secondly, spatial normalization was performed to co-register the different images into the common coordinate space MNI152; an affine registration, with default configuration, was performed with the FSL Linear Registration Tool (FLIRT). Linear registration is preferred to a non-linear for the methodology to be robust to subtle local differences, due for example to subject morphological variability, or small registration failures. Finally, using the template brain coordinates, we automatically segmented the brain of each subject into the two hemispheres and, starting from the medial longitudinal fissure plane, uniformly divided each hemisphere in an equal number of rectangular ($l_1 \times l_2 \times l_3$) patches, covering the whole brain, see Figure \ref{fig:figure_patches}. Only those patches overlapping with the template brain for more than $10\%$ were kept.

Patches were the nodes of a network whose connections represented the grade of similarity between them. Several similarity metrics were explored (see the ``Similarity Metric Study'' section in Supplementary Materials) and Pearson's correlation was finally preferred. 

For each subject, we built an undirected weighted network with edges defined by pairwise Pearson's correlation among patches. Therefore, multiplex network $\pmb{\mathcal{G}} = \{\mathcal{G}_1,\mathcal{G}_2,...,\mathcal{G}_\alpha,..., \mathcal{G}_M\}$ was, in this case, a collection of single subject weighted networks  $\mathcal{G}_\alpha=(N,\mathcal{E}_\alpha,\mathcal{W}_\alpha)$ sharing a common number of nodes $N$, while the set of links $\mathcal{E}_\alpha$ and the inherent weights $w_{ij} \in \mathcal{W}_\alpha$ change depending on the layer $\alpha$. According to this notation, each network $\mathcal{G}_\alpha$ can also be represented by the corresponding adjacency matrix $A_\alpha = {a^\alpha_{ij}}$.

Pearson's correlation admits continuous values in the $[-1,1]$ interval. Negative correlations were disregarded. It is worth noting that negative correlations can be found, for example, between patches in which gray matter and white matter undergo a left-right inversion. As a result, distinguishing positive and negative correlations would include in the multiplex model a left-right bias. As asymmetry is a common characteristic of atrophy in AD, it was decided to consider undirected networks. In addition, we decided to threshold the networks by setting to $0$ all connections whose absolute correlation was less than moderate ($|r|<.3$), in order to exclude noisy interrelationships in the model, and reducing as much as possible the loss of important links. For higher correlations, weights were kept in the model, thus resulting in a weighted undirected network representation for each subject. An investigation on how the threshold affects the multiplex network ability to detect diseased patterns is also reported in the ``Threshold Study'' section in Supplementary Materials.

%

%
%

\subsection*{Multiplex features}
\label{sec:feats}

In a multiplex it is possible to introduce several topological characteristics that are usually adopted to describe a complex network \cite{menichetti2014weighted}. In our approach we employed the following indicators: the strength $s_i^\alpha$  and the inverse participation ratio $Y_i^\alpha$ of a node $i$ in layer $\alpha$:

\begin{equation}
s_i^\alpha = \sum_{j=1}^N w_{ij}^\alpha
\end{equation}

\begin{equation}
Y_i^\alpha = \sum_{j=1}^N \left(\frac{w_{ij}^\alpha}{s_i^\alpha}\right)^2
\end{equation}

\noindent Strength measurements denote which nodes are more relevant within the network describing a single layer (\textit{i. e.} a subject) of the multiplex. Inverse participation ratio attains the heterogeneity of the weight distribution within each layer.

Along with these two measurements we also evaluated the conditional means of strength $s(k)^\alpha$ and inverse participation $Y(k)^\alpha$ against the nodes with degree $k$:

\begin{equation}
s(k)^\alpha = \frac{1}{N_k} \sum_{i =1}^N s_{i}^\alpha \delta(k_i^\alpha,k)
\end{equation}

\begin{equation}
Y(k)^\alpha = \frac{1}{N_k} \sum_{i =1}^N Y_{i}^\alpha \delta(k_i^\alpha,k)
\end{equation}

\noindent
Summation is extended over the $N_k$ nodes having degree $k$; as summation includes a Kronecker $\delta$ function, the only non-null terms, for both strength and inverse participation, are referred to nodes $i$ of the layer $\alpha$ whose degree is $k$. These quantities help to understand how weights are distributed within each layer, thus, for example, distinguishing whether, on average, the weights of central nodes and less connected nodes are identically distributed or not. Several studies have already pointed out, especially with group-wise single layer approaches \cite{tijms2013single}, how these features can describe significant differences among healthy and diseased subjects. 

However, it is reasonable to assume that further evidence of significant differences between subjects, can arise from the context information provided by the multiplex framework. Accordingly, this information content was exploited by considering the aggregate adjacency matrix $A^{multi}={a}^{multi}_{ij}$ where:

\begin{equation}
a^{multi}_{ij}= \{1\ if\ \exists \alpha | w_{ij}^\alpha >0\quad \wedge \quad 0 \quad otherwise\}
\end{equation}

\noindent The matrix $A^{multi}$ naturally allowed us to re-introduce the previous measurements within a global perspective. In fact, it was possible to compute for each node an aggregated degree and then use it to weight the previously defined strength and inverse participation. Analogously, we used $A^{multi}$ to define the aggregate degree for each node and then re-computing the conditional means. In this way we introduced in the description of each node the information produced by the whole multiplex. 

In conclusion each network was described by $8N$ features ($4N$ single layer and $4N$ multiplex features), resulting in a $M \times 8N$ feature representation. It is worthwhile to note that this characterization was independent from the clinical status of the subjects as the multiplex had been built blindly to diagnosis. This base of knowledge was then investigated with supervised machine learning models to extract specific disease effect patterns.

\subsection*{Classification and Clinical feature importance}
\label{sec:class}

\begin{figure}[!htbp]
\centering
\includegraphics[scale=.5]{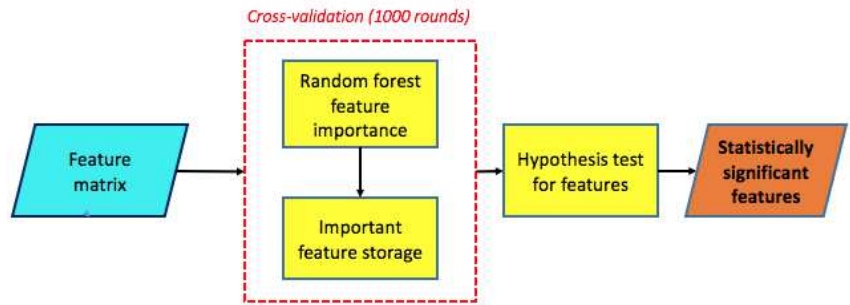} 
\caption{A flowchart of the feature selection methodology: the features, stored in a matrix, are used to train a random forest model, this model provides a feature important estimation; the procedure is cross-validated with a $5$-fold for $1000$ times, at each round taking into account the selected feature. Finally, a statistical test of hypothesis establishes which features have been selected a significant number of times.} 
\label{fig:feat_selec}
\end{figure}

The multiplex characterization of the images yielded a simple matrix representation, which could be used to feed machine learning models, and unveil discriminating anatomical patterns. The number of features $f$, involved in this approach, could easily reach values ranging from $\sim10^3$ to $\sim10^4$ outnumbering the number of the available training samples. Thus, to prevent over-training issues, arising from the curse of dimensionality and assess the multiplex framework, a feature selection was necessary. A flowchart of the whole feature selection method is represented in Figure \ref{fig:feat_selec}.

A $5$-fold cross-validation feature importance selection was performed within a wrapper-based strategy. We randomly divided $1000$ times $\mathcal{A}_{train}$ in a training and a validation test. For each cross-validation round we built a multiplex model on training subjects, then we computed the related training multiplex features and the overall matrix $\mathcal{A}_{train}$. For test subjects, single layer features were straightforwardly computed. Features accounting the whole multiplex structure were in turn computed adding the test subject to the training multiplex but keeping fixed $\mathcal{A}_{train}$. The reason for this choice can be justified considering the perturbation induced by the addition of one layer is small.

For each cross-validation round, we trained a first random forest classifier and selected the most important features. In particular, we measured the total decrease in node impurities, in terms of Gini index, from splitting on the variable, averaged over all trees. The selected features were stored for later use and used to train a second random forest classifier which was used to predict the diagnosis of the validation subjects. In both cases random forests were grown with $500$ trees, a number large enough for the out-of-bag error to reach the typical training plateau. 

At each split $\sqrt{f}$ features were randomly sampled, thus, for each cross-validation round, different features were selected; to determine the most important features, we measured the overall occurrence rate of each feature, interpreting it as a success rate. As a consequence, we compared the probability to observe such occurrence with a binomial distribution and an experimental p-value could be computed to test the randomness hypothesis. To select a more exiguous number of features a p-value $< 0.01$ threshold was set, then we established which ones had shown a significant probability of occurrence. Once the best features had been selected, we used them to train a new ensemble model on $\mathcal{D}_{train}$ and tested it on $\mathcal{D}_{test}$ to assess the method robustness and evaluate the informative content carried by multiplex features.

It is worth noting that features like strength and inverse participation have a direct interpretation, being directly related to a single patch of the brain network whilst conditional means, by definition, are related to several nodes sharing a common degree $k$. For classification purposes this is not an issue, being based on computed features; on the contrary this is relevant in order to provide an anatomical interpretation and a diagnostic value of the features selected. Again, for each significant feature we determined with the same strategy which anatomical districts had show a significant association to AD.


\section*{Supplementary Material}

\subsection*{Demographic Information}
\label{sec:dem}

Clinical and demographic information, including the Mini Mental State Examination (MMSE) score, age, years of education and sex for the $\mathcal{D}_{train}$ and $\mathcal{D}_{test}$ is detailed in Table \ref{tab_db}. Except for MMSE scores, there were no significant differences among the three groups according to a Wilcoxon test.

\begin{table}[!htbp]
\centering
\begin{tabular}{|l|ll|lll|l|}
\hline
       							&   $\mathbf{\mathcal{D}_{train}}$  &                       			&  $\mathbf{\mathcal{D}_{test}}$ 	&          					&   					& \textbf{Total} 				\\
\hline
Disease status						&   AD (38) 					&   	  NC (29) 			&   AD (48) 					&     NC (52) 				&	cMCI (48)			& $215$	    			\\
Female/male					&	$18/20$ 					& 		 $13/16$			&	$22/26$ 					&     $25/27$			&   $21/27$   		& $99/116$			\\
Age (years)					&	$74\ \pm\ 8$ 			& 		 $75\ \pm\ 6$	&	$78\ \pm\ 6$ 			& 	  $75\ \pm\ 6$		&   $76\ \pm\ 6$   	& $76\ \pm\ 6$		\\
Education (years)			&	$15\ \pm\ 3$ 			& 		 $17\ \pm\ 3$	&	$15\ \pm\ 3$ 			&     $16\ \pm\ 3$		&   $15\ \pm\ 3$   	& $16\ \pm\ 3$		\\
MMSE 							&	$23\ \pm\ 2$ 			& 		 $29\ \pm\ 1$	&	$24\ \pm\ 2$ 			& 	  $29\ \pm\ 1$		&   $27\ \pm\ 2$   	& $26\ \pm\ 2$		\\
n sites 						&	$23$ 						& 		 $19$				&	$26$ 						& 	  $29$					&   $18$   			& $34$				\\
\hline
\end{tabular}
\caption{Group size and sex information are reported for each class. The table also provides age, years of education and MMSE (mean and standard deviation). The disease status reported is as assessed at baseline visit. MMSE score resulted statistically different for all groups with a p-value $< 0.01$ except between $\mathcal{D}_{train}$ and $\mathcal{D}_{test}$ normal controls and between $\mathcal{D}_{train}$ and $\mathcal{D}_{test}$ AD patients.}
\label{tab_db}
\end{table}

\subsection*{Threshold Study}
\label{sec:thr}

\begin{figure}[!htbp]
\centering
\includegraphics[angle=270,scale=.28]{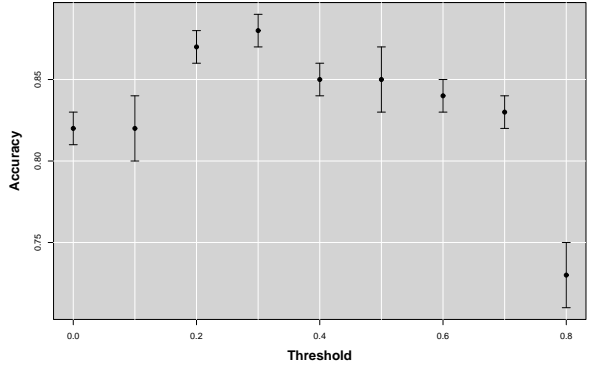} 
\caption{The figure shows the accuracy as a function of the threshold that changes from $0$ to $0.8$. The best accuracy is obtained in correspondence of a threshold value of $0.3$. }
\label{fig:thr_study}
\end{figure}

Since this approach could in principle heavily depend on the threshold value adopted to discard negligible correlations, the threshold values ranging from $0$ to $0.8$ were explored with a $0.1$ step. Then, for each threshold value a different multiplex was constructed. The patch dimension adopted was $3000$ mm\textsuperscript{3}. The training classification performance was measured in terms of accuracy, see Figure \ref{fig:thr_study}

The classification accuracy reached its maximum value with a $0.3$ threshold value and it remained stable over $0.85$ for a large range of correlations $[0.2,0.5]$. Outside this range a performance drop was observed. With lower or higher threshold performances showed a significant decrease, especially above the $0.8$ threshold; in which case more of the $50\%$ of the networks resulted empty. This is because lower threshold values introduced noisy correlations within the model, thus concealing the effective network information, whilst greater threshold values were too penalizing as informative links were neglected.

\subsection*{Similarity Metric Study}
\label{sec:sim_metric}

We investigated different similarity measures \cite{razlighi2013evaluating} to define the presence of an edge between two generic patches $s_i$ and $s_j$. In addition to the Pearson's correlation (PC), we studied:
\begin{itemize}
\item \textit{Mutual Information} (MI): 

\begin{equation*}
MI_{ij}= H(s_i) + H(s_j) - H(s_i,s_j) 
\end{equation*}

\noindent 
where $H(s_i)$ and $H(s_j)$ are the Shannon entropies related to the patches $s_i$, $s_j$ and $H(s_i,s_j)$ is their joint Shannon entropy. 
 
\item \textit{Mean Square Differences} (MSD): 
\begin{equation*}
MSD_{ij} = \frac{1}{D} \sum_{k=1}^D (s_i^k-s_j^k)^2 
\end{equation*}

\noindent 
with $s_i^k$, $s_j^k$ being the voxel intensity within a patch and $D$ the total number of voxels.

\item \textit{Hellinger distance} (HD): 

\begin{equation*}
HD_{ij} = \frac{1}{D} \sqrt{\sum_{k=1}^D (\sqrt{s_i^k}-\sqrt{s_j^k})^2} 
\end{equation*}

\noindent 
as usual $D$ is the patch size and $s_i^k$, $s_j^k$ are the voxel intensities.

\item \textit{Kolmogorov Smirnov non parametric statistic test} (KST) quantifying the shape difference between gray level distributions of the patch pairs.
\item \textit{Unpaired t statistic test} (UtT) evaluating the difference between means of the patch gray level distribution pairs in terms of standard error.
\end{itemize} 

For MI the same $0.3$ threshold used for Pearson's Correlation was adopted. For MSD and HD, adjacency matrix was obtained computing the complementary of the normalized MSD and HD matrix and placing to zero the values $<0.3$. Finally for KS and unpaired t test, adjacency matrix was given by the complementary of the normalized test statistic matrix, putting to zero the elements rejecting respectively the null hypothesis of patch distribution and mean equality with a p-value $<0.05$. A complete summary of the metric study, including sensitivity and specificity, is reported in Table \ref{tab_metric}: 

%

\begin{table}[!htbp]
\centering
\begin{tabular}{|l|l|l|l|l|}

\hline
       	\textbf{Similarity Metric}						&   \textbf{Accuracy} &                       			 \textbf{Sensitivity} 	& \textbf{Specificity}	\\
\hline
Pearson's Correlation & $\mathbf{0.88 \pm 0.01}$ & $\mathbf{0.90 \pm 0.01}$ & $\mathbf{0.88 \pm 0.02}$ \\
\hline
Mutual Information & $0.87 \pm 0.01$ & $\mathbf{0.90 \pm 0.01}$ & $0.87 \pm 0.02$ \\
\hline
Mean Square Differences & $0.81 \pm 0.01$ & $0.84 \pm 0.01$ & $0.80 \pm 0.02$\\ 
\hline
Hellinger Distances & $0.80 \pm 0.01$ & $0.83 \pm 0.01$ & $0.77 \pm 0.02$\\
\hline
Kolmogorov Smirnov test & $0.77 \pm 0.01$ & $0.82 \pm 0.01$ & $0.77\pm  0.02$ \\ 
\hline
Unpaired t test & $0.75 \pm 0.01$ & $0.79 \pm 0.01$ & $0.73 \pm 0.02$ \\
\hline
\end{tabular}
\caption{For each similarity measurement, accuracy sensitivity and specificity with relative standard errors are shown. Best performing metrics are indicated in bold.}
\label{tab_metric}
\end{table}

PC and MI are intrinsically normalized metrics which also exploit the spatial correspondence of voxels within a patch; they gave best results. MSD  and HD, lacking normalization, suffer a significant performance drop. KST and UtT consider respectively only the shape and the average of the patch gray level distribution gave lower performances.

\bibliography{biblio}


\section*{Acknowledgements }

Data used in the preparation of this article was obtained from the ADNI database (adni.loni.usc.edu). The ADNI was launched in 2003 by the National Institute on Aging (NIA), the National Institute of Biomedical Imaging and Bioengineering (NIBIB), the Food and Drug Administration (FDA), private pharmaceutical companies and non-profit organizations, as a 60 million, 5 year public-private partnership. The primary goal of ADNI has been to test whether serial MRI, positron emission tomography (PET), other biological markers, and clinical and neuropsychological assessment can be combined to measure the progression of mild cognitive impairment (MCI) and early Alzheimer’s disease (AD). Determination of sensitive and specific markers of very early AD progression is intended to aid researchers and clinicians to develop new treatments and monitor their effectiveness, as well as lessen the time and cost of clinical trials. The Principal Investigator of this initiative is M W Weiner, MD, VA Medical Center and University of California—San Francisco. ADNI is the result of efforts of many coinvestigators from a broad range of academic institutions and private corporations, and subjects have been recruited from over 50 sites across the U.S. and Canada. The initial goal of ADNI was to recruit 800 subjects but ADNI has been followed by ADNI-GO and ADNI-2. To date these three protocols have recruited over 1500 adults, ages 55 to 90, to participate in the research, consisting of cognitively normal older individuals, people with early or late MCI, and people with early AD. The follow up duration of each group is specified in the protocols for ADNI-1, ADNI-2 and ADNI-GO. Subjects originally recruited for ADNI-1 and ADNI-GO had the option to be followed in ADNI-2. For up-to-date information, see www.adni-info.org. Data collection and sharing for this project was funded by the Alzheimer’s Disease Neuroimaging Initiative (ADNI) (National Institutes of Health Grant U01 AG024904) and DOD ADNI (Department of Defense award number W81XWH-12-2-0012). ADNI is funded by the National Institute on Aging, the National Institute of Biomedical Imaging and Bioengineering, and through generous contributions from the following: Alzheimer’s Association; Alzheimer’s Drug Discovery Foundation; BioClinica, Inc.; Biogen Idec Inc.; Bristol-Myers Squibb Company; Eisai Inc.; Elan Pharmaceuticals, Inc.; Eli Lilly and Company; F Hoffmann-La Roche Ltd and its affiliated company Genentech, Inc.; GE Healthcare; Innogenetics, N V; IXICO Ltd.; Janssen Alzheimer Immunotherapy Research \& Development, LLC.; Johnson \& Johnson Pharmaceutical Research \& Development LLC.; Medpace, Inc.; Merck \& Co., Inc.; Meso Scale Diagnostics, LLC.; NeuroRx Research; Novartis Pharmaceuticals Corporation; Pfizer Inc.; Piramal Imaging; Servier; Synarc Inc.; and Takeda Pharmaceutical Company. The Canadian Institutes of Health Research is providing funds to support ADNI clinical sites in Canada. Private sector contributions are facilitated by the Foundation for the National Institutes of Health (www.fnih.org). The grantee organization is the Northern California Institute for Research and Education, and the study is coordinated by the Alzheimer’s Disease Cooperative Study at the University of California, San Diego. ADNI data are disseminated by the Laboratory for Neuro Imaging at the University of Southern California. All authors disclose any actual or potential conflicts of interest, including any financial, personal, or other relationships with other people or organizations that could inappropriately influence their work. All experiments were performed with the informed consent of each participant or caregiver in line with the Code of Ethics of the World Medical Association (Declaration of Helsinki). Local institutional ethics committees approved the study.

\section*{Author contributions statement}


N.A. and M.L. conceived and conducted the analyses, S.B gave clinical support. All authors analysed the results and reviewed the manuscript. 

\section*{Additional information}


The corresponding author is responsible for submitting a \href{http://www.nature.com/srep/policies/index.html#competing}{competing financial interests statement} on behalf of all authors of the paper. This statement must be included in the submitted article file.

\end{document}